\documentclass[a4paper]{jpconf}
\usepackage{graphicx}
\usepackage{mathrsfs}
\begin{document}
\title{Single crystal $^{27}$Al-NMR study of the cubic $\Gamma$$_3$ ground doublet system PrTi$_2$Al$_{20}$}

\author{T Taniguchi, M Yoshida, H Takeda, M Takigawa, M Tsujimoto, A Sakai, Y Matsumoto, and S Nakatsuji}

\address{Institute for Solid States Physics, University of Tokyo, Kashiwa 277-8581, Japan}

\ead{taka.taniguchi@issp.u-tokyo.ac.jp}

\begin{abstract}
We report results of $^{27}$Al-NMR measurements on a single crystal of PrTi$_2$Al$_{20}$, in which the ground state of Pr$^{3+}$ ions in the crystalline electric field is the nonmagneitc $\Gamma$$_3$ doublet. From the analysis of NMR spectra with the magnetic field applied precisely along the $\left\langle 111\right\rangle $ and $\left\langle 100\right\rangle$ directions, we determined the electric field gradient tensors for all three inequivalent Al sites (Al(1) $\sim$ Al(3) sites) and the anisotropic Knight shifts for the Al(3) sites. The hyperfine coupling tensor at the Al(3) sites is strongly anisotropic and much larger than the classical dipolar coupling, indicating importance of the anisotropic hybridization between the conduction and $f$ electron states. 
\end{abstract}

\section{Introduction}

Recently, $f$ electron compounds with a nonmagnetic ground state of the crystalline electric field (CEF) potential have attracted much interest. A number of non-trivial phenomena such as multipole order or quadrupole Kondo effects are expected at low temperatures, when the nonmagnetic CEF ground state has non-zero matrix elements of the quadrupole and octupole momoents [1,2]. Particularly interesting examples are the quadrupole order and unconventional superconductivity reported on Pr$T_2$$X_{20}$ ($T$:transition metals, $X$:Zn, Al, $\cdots$), where non-Kramers Pr$^{3+}$ ions (4$f^2$ system) are encapsulated in the Frank-Kasper cage formed by $X$ atoms [3-8]. In a cubic CEF potential, the 4$f^2$ ground multiplet of $J$=4 splits into the $\Gamma_1$ nonmagnetic singlet,  $\Gamma_3$ nonmagnetic doublet,  $\Gamma_4$ magnetic triplet, and $\Gamma_5$ magnetic triplet. 

Figure 1 shows the temperature dependence of the magnetic susceptibility $M$/$H$ of PrTi$_2$Al$_{20}$ at 6.615 T for $H$ $\parallel$ $\left\langle 111\right\rangle $. $M$/$H$ is nearly independent of temperature below 20 K, indicating a nonmagnetic CEF ground state with Van Vleck paramagnetism. The specific heat measurements show that the 4$f$ electronic entropy reaches $R$ln2 near 5 K, suggesting that the ground state is the $\Gamma_3$ doublet [5]. $M$/$H$ increases with decreasing temperature above 20 K due to the excited $\Gamma_4$ and $\Gamma_5$ triplets. This result is consistent with the CEF level scheme deduced from the inelastic neutron scattering experiments, which determined the energies between the ground $\Gamma_3$ and the excited $\Gamma_4$, $\Gamma_5$, and $\Gamma_1$ levels to be 65, 108, and 156~K, respectively [9]. A phase transition was detected by an anomaly near 2~K in the specific heat [5]. A clear anomaly in the ultrasonic measurement combined with the absence of anomaly in the magnetic susceptibility indicates that it is a quadrupole transition [10]. At ambient pressure, the superconductivity appears below $T_c$ =200 mK. Application of pressure above 6 GPa causes decrease of the quadrupole transition temperature and increase of the superconducting $T_c$, which reaches near 1.1 K at 8.7 GPa [8]. The electron effective mass deduced from the upper critical field is also enhanced by pressure exceeding 100 times the bare electron mass [11]. This raises a possibility that the quadrupole fluctuations play a vital role for the formation of Cooper pairs.

\begin{figure}[h]
\centering 
\begin{minipage}{17pc}
\includegraphics[width=18pc]{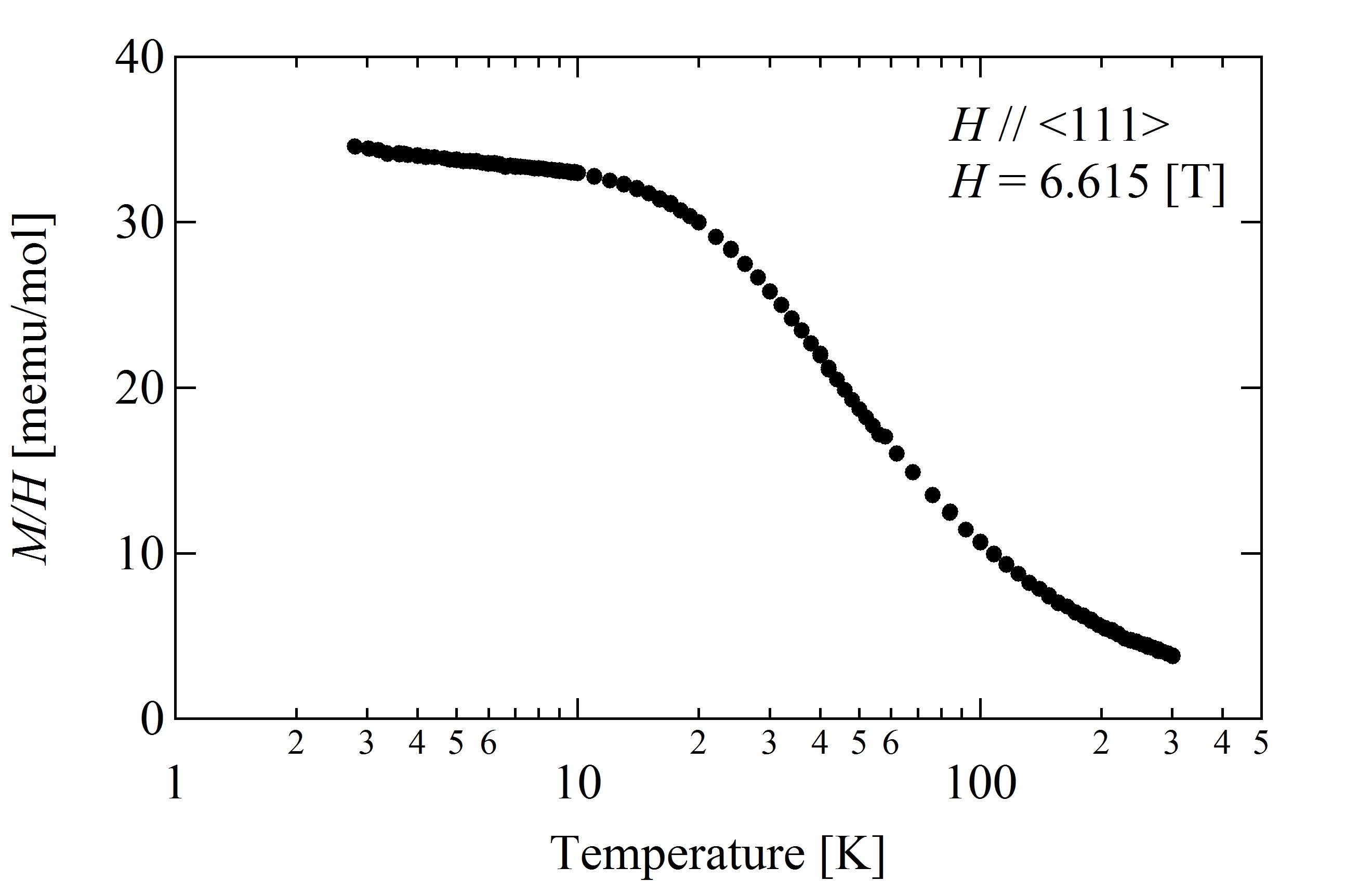}
\caption{\label{label} 
The temperature dependence of the magnetic susceptibility of PrTi$_2$Al$_{20}$ at 6.615~T for $H$ $\parallel$ $\left\langle 111\right\rangle $
}
\end{minipage}\hspace{2pc}%
\begin{minipage}{17pc}
\includegraphics[width=15pc]{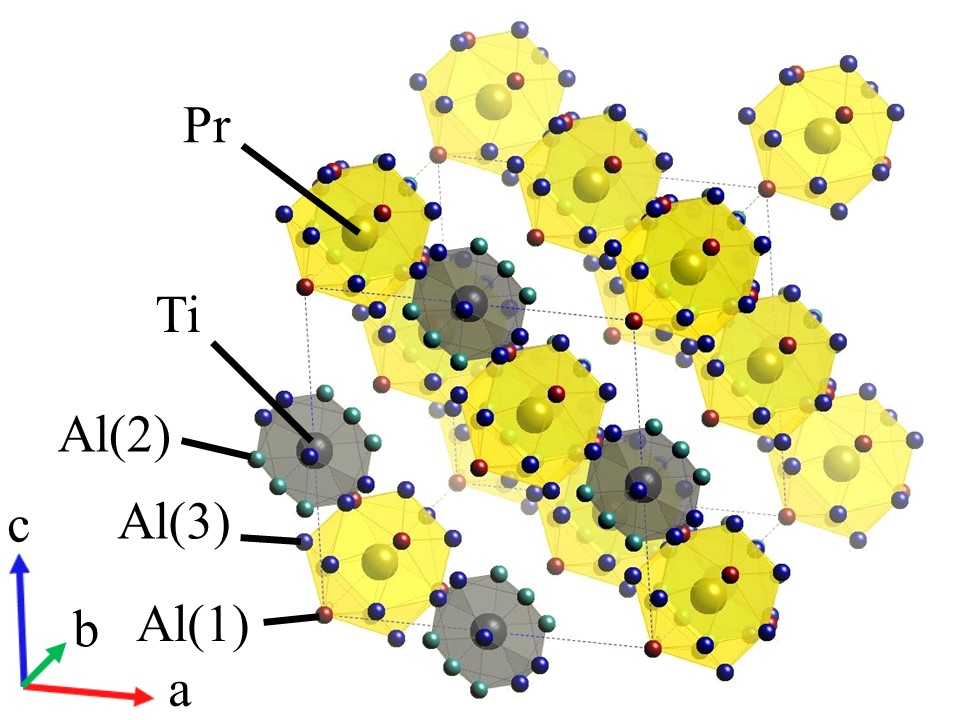}
\caption{\label{label}
The crystal structure of PrTi$_2$Al$_{20}$. The Pr and Ti atoms are surrounded by yellow and gray cages, respectivity.
}
\end{minipage} 
\end{figure}

Figure 2 shows the crystal structure of PrTi$_2$Al$_{20}$. This compound has the cubic CeCr$_2$Al$_{20}$ type structure with the lattice parameter $a$=14.723 \AA and the space group $Fd\bar{3}m$ [12]. There are three crystallographycally inequivalent Al sites, Al(1), Al(2), and Al(3), occupying the 16$c$, 48$f$ and 96$g$ Wyckoff positions with the point symmetries $\bar{3}m$, $2mm$, and $m$, respectively. Pr atoms are surrounded by four Al(1) atoms and twelve Al(3) atoms. Ti atoms are surrounded by six Al(2) atoms and six Al(3) atoms. 

Tokunaga \textit{et al.} reported the results of NMR Knight shift and nuclear magnetic relaxation rates for selected Al sites and field orientations [13]. In this paper, we report measurements of the electric field gradient (EFG) and the Knight shift for all Al sites. In particular, we have completely determined the EFG tensor for all Al sites and the magnetic hyperfine coupling tensor between Pr-4$f$ electrons and $^{27}$Al nuclei at the Al(3) sites. 

\section{Experimental procedure}
A single crystal of PrTi$_2$Al$_{20}$ was synthesized by the Al flux method [5]. For NMR measurements, the crystal was shaped into a thin plate of the size 2.05 $\times$ 1.14 $\times$ 0.073 mm$^3$ with the $\left\langle 111\right\rangle $ direction normal to the plate. There are two reasons for this. First, thinner plates give better signal intensity because the rf penetration depth is much shorter than the thickness. Second, the geometry of a thin plate reduces the distribution of demagnetizing field, thereby resulting in a narrower NMR line width. The NMR spectra were obtained by summing the Fourier transform of the spin-echo signals obtained at equally spaced rf frequencies with a fixed magnetic field of 6.615 T. The orientation of the crystal in the magnetic field was precisely controlled by double-axis goniometer typically within 0.2 degree.

\section{NMR spectra and EFG}
In general, the NMR resonance frequency is determined by the spacing of eigenvalues of the following Hamiltonian
\begin{equation}
\mathscr{H}=-\gamma\hbar\mathbf{I}\cdot\mathbf{H}+\frac{1}{6}h\nu_{Q}\left\{ \left(3I_{Z}^{2}-I^{2}\right)+\eta\left(I_{X}^{2}-I_{Y}^{2}\right)\right\} .
\end{equation}
The first term represents the Zeeman interaction, where $h$ is the Plank constant, $\hbar=h/2\pi$, $\mathbf{I}$ is the nuclear spin and $\gamma$ is the gyromagnetic ratio (11.09407~MHz/T for $^{27}$Al). The effective magnetic field $\mathbf{H}$ acting on nuclei is related to the external magnetic field $\mathbf{H}_{\rm{ext}}$ as $\mathbf{H}=\mathbf{H}_{\rm{ext}}\mathrm{\left(1+\mathit{K}\right)}$, where $K$ is the Knight shift. The second term corresponds to the nuclear quadrupole interaction. The quadrupole coupling constant $\nu_{Q}$ is defined as $\nu_{Q}=\frac{3eQ}{2I\left(2I-1\right)}V_{ZZ}$, where $e$ is the proton's charge, $Q$ is the nuclear quadrupole moment ($0.149\times10^{-24}$ cm$^{-2}$ for  $^{27}$Al) and $V_{ZZ}=\frac{\partial^{2}V}{\partial Z^{2}}$ is the largest principal value of the EFG tensor at the nuclei. Here $X$, $Y$, and $Z$ denote the principal axes of the EFG tensor, satisfying the relation $\left|V_{ZZ}\right|\geq\left|V_{YY}\right|\geq\left|V_{XX}\right|$ and the symmetry parameter of EFG is defined as $\eta=\frac{V_{YY}-V_{XX}}{V_{ZZ}}$ ($0\leq\eta\leq1$, note that $V_{XX}+V_{YY}+V_{ZZ}=0$).

\begin{figure}
\begin{center}
\includegraphics[width=35pc]{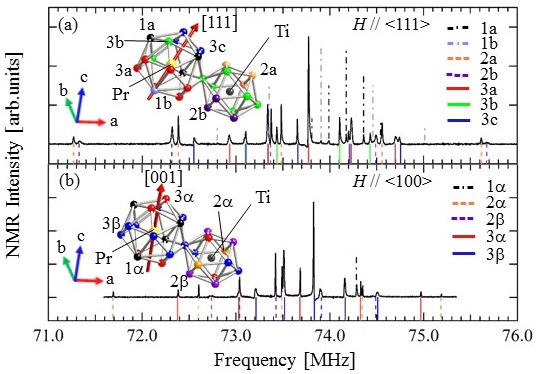}
\end{center}
\caption{\label{label}
NMR spectra of $^{27}$Al nuclei at 6.615 T and 60 K for (a) $\mathbf{H}_{\rm{ext}} \parallel \left\langle 111\right\rangle $ and (b) $\mathbf{H}_{\rm{ext}} \parallel \left\langle 100\right\rangle $. The vertical lines show the results of fitting by equation (1) as described in the text. The insets show how the Al(1), Al(2), and Al(3) sites split into inequivalent groups of sites in magnetic fields. The colors of the lines indicate the site assignment. 
}
\end{figure}

Figure 3 shows the $^{27}$Al-NMR spectra at 6.615 T and 60 K for (a) $\mathbf{H}_{\rm{ext}} \parallel \left\langle 111\right\rangle $ and (b) $\mathbf{H}_{\rm{ext}} \parallel \left\langle 100\right\rangle $. The observed spectra consist of sharp NMR lines, confirming high quality of the single crystal. Since $I=5/2$ for $^{27}$Al nuclei, NMR spectrum of each site should consist of five resonance lines split by the quadrupole interaction. Furthermore, crystallographically equivalent sites generally split into several inequivalent groups of sites in magnetic fields. We have fitted the frequencies of all the observed resonance lines in figure 3 to the calculated values obtained by diagonalizing the Hamiltonian of equation (1) in order to determine the values of $K$, $\nu_{Q}$, $\eta$, and the directions of the principal axes of EFG as described in detail below. This process also allowed us to make unique site assignment of observed resonance lines as indicated in figure 3.  

\subsection*{\underline{Al(1)site}}
The Al(1) site has a 3-fold symmetry along the $\left\langle 111\right\rangle $  direction. Therefore the EFG tensor is axially symmetric ($\eta = 0$) and the principal axis with the largest EFG ($Z$ axis) is parallel to the $\left\langle 111\right\rangle $  direction. There are four equivalent $\left\langle 111\right\rangle $ directions in the cubic crystal. Thus, when $\mathbf{H}_{\rm{ext}} \parallel \left\langle 111\right\rangle $, the $Z$ axis is parallel to the field for one quarter of A(1) sites while it is not for the other three quarter of sites. Therefore, Al(1) sites split into two sites with the ratio of 3:1 denoted 1a and 1b in figure 3a and the spectrum form the Al(1) sites consists of ten resonance lines. When $\mathbf{H}_{\rm{ext}} \parallel \left\langle 100\right\rangle $, all the $\left\langle 111\right\rangle $ directions make an identical angle of 54.7 degree with the field, therefore all the Al(1) sites become equivalent (denoted 1$\alpha$ in figure 3b). Furthermore, this is the \textit{magic angle}, at which quadrupole splitting vanishes to the first order in the quadrupole interaction. Thus the spectrum form the Al(1) site consists of a single line. By fitting the experimental frequencies of the eleven resonance lines to the calculated values obtained by diagonalizing equation (1), we have determined the four parameters: $\nu_{Q}=0.55\pm0.02$ MHz, $K\left(1a\right)=1.073\pm0.005$\%, $K\left(1b\right)=0.707\pm0.003$\%, and $K\left(1\alpha\right)=1.221\pm0.011$\%.

\subsection*{\underline{Al(2)site}}
Figure 4 shows a part of the crystal structure, visualizing the $2mm$ point symmetry of the Al(2) sites. The Al(2) site marked by the arrow is located on the two mirror planes, $\left(0\bar{1}1\right)$ and $\left(011\right)$. Therefore, two of the principal axes of EFG should be perpendicular to these planes, $\left[0\bar{1}1\right]$ and $\left[011\right]$. The third one then must be along $\left[100\right]$, which is the intersection of the two mirror planes. The local symmetry, however, does not determine which one corresponds to the largest principal value of EFG. For $\mathbf{H}_{\rm{ext}} \parallel \left[111\right]$, the Al(2) sites split into two inequivalent sites with 1:1 ratio denoted 2a and 2b in figure 4 (the red and blue circles). For $\mathbf{H}_{\rm{ext}} \parallel \left[001\right]$, there are also two inequivalent sites but with 1:2 ratio denoted 2$\alpha$ and 2$\beta$ (the open and closed circles). In each case, the NMR spectrum of the Al(2) sites consists of ten resonance lines. From the observed frequencies of these lines, we have determined six parameters: $\nu_{Q}=2.07\pm0.01$ MHz, $\eta=0.155\pm0.001$, $K\left(2a\right)=0.091\pm0.003$\%, $K\left(2b\right)=0.038\pm0.005$\%, $K\left(2\alpha\right)=0.097\pm0.005$\%, and $K\left(2\beta\right)=0.130\pm0.008$\%. In order to reproduce the experimental results, the $Z$ ($X$) axis corresponding to the largest (smallest) principal value of EFG must be assigned to the $\left[011\right]$ ($\left[100\right]$) direction for the Al(2) site marked in figure 4.

\subsection*{\underline{Al(3)site}}
Figure 5 shows a part of the crystal structure, visualizing the $m$ point symmetry of the Al(3) sites that form a cage surrounding a Pr atom. The Al(3) site indicated by the arrow is on the $\left(1\bar{1}0\right)$ mirror plane. Therefore one of the principle axes of the EFG should be along the $\left[1\bar{1}0\right]$ direction perpendicular to the mirror plane. Other two principle axes are in the mirror plane. We define $\alpha$ to be the angle between one of them and the $\left[110\right]$ direction. For  $\mathbf{H}_{\rm{ext}} \parallel \left[111\right]$, the Al(3) sites split into three inequivalent sites with 2:1:1 ratio denoted 3a, 3b and 3c sites in figure 5 (the red, green, and blue circles, respectively), generating fifteen resonance lines. For  $\mathbf{H}_{\rm{ext}} \parallel \left[001\right]$, the Al(3) sites split into two inequivalent sites with 1:2 ratio denoted 3$\alpha$ and 3$\beta$ (open and closed circles), generating ten resonance lines. From the observed twenty five lines, we have determined eight parameters: $\alpha=18.1\pm1.8$ degree, $\nu_{Q}=0.99\pm0.01$ MHz, $\eta=0.594\pm0.027$, $K\left(3a\right)=0.558\pm0.001$\%, $K\left(3b\right)=0.519\pm0.005$\%, $K\left(3c\right)=0.357\pm0.012$\%, $K\left(3\alpha\right)=0.399\pm0.008$\%, and $K\left(3\beta\right)=0.608\pm0.021$\%. In order to reproduce the experimental results, the $Z$ axis corresponding to the largest principal value of EFG must be assigned to the the $\left[1\bar{1}0\right]$ direction and the $X$ axis should be tilted from $\left[110\right]$ direction by $\alpha$ as shown in figure 5. 

\subsection*{}
The obtained values of the EFG parameters are summarized in table 1. These are in agreement with the previous results [13] except for minor discrepancies. Our results are based on larger number of data points and should be considered as the refinement. 

\begin{center}
\begin{table}[h]
\caption{EFG parameters obtained from the NMR spectra at 60 K.}
\centering
\begin{tabular}{@{}*{7}{c}}
\br
site&$\nu_{Q}$ [MHz]&$\eta$&$\alpha$ [deg.]\\
\mr
Al(1)&$0.55\pm0.02$&0&-\\
Al(2)&$2.07\pm0.01$&$0.155\pm0.001$&-\\
Al(3)&$0.99\pm0.01$&$0.594\pm0.027$&$18.1\pm1.8$\\
\br
\end{tabular}
\end{table}
\end{center}

\begin{figure}[h]
\centering 
\begin{minipage}{17pc}
\includegraphics[width=14pc]{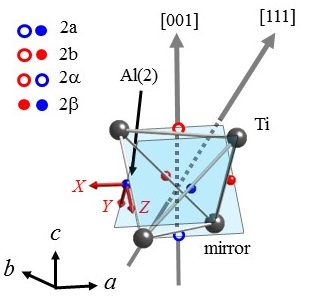}
\caption{\label{label}
A part of the crystal structure showing the point symmetry the Al(2) site. The semitransparent sheets represent the two mirror planes of the Al(2) sites marked by the black arrow. The principal axes of EFG of this site  ($X$, $Y$, and $Z$) are indicated by the red arrows, with $Z$ ($X$) axis corresponding to the largest (smallest) principal value of EFG. The magnetic field along the $\left[111\right]$ direction generates two inequivalent sites denoted 2a and 2b (the red and blue circles), while the field along the $\left[001\right]$ direction generates different set of inequivalent sites denoted 2$\alpha$ and 2$\beta$ (the open and close circles).
}
\end{minipage}\hspace{2pc}%
\begin{minipage}{17pc}
\includegraphics[width=15pc]{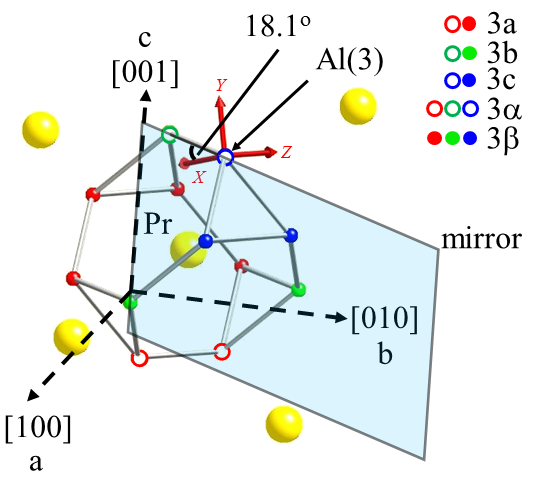}
\caption{\label{label}
A part of the crystal structure showing the point symmetry of the Al(3) site. The  semitransparent sheet represents the mirror plane of the Al(3) site marked by the black arrow. The principal axes of EFG ($X$, $Y$, and $Z$) are indicated by the red arrows, with $Z$ ($X$) axis corresponding to the largest (smallest) principal value of EFG. The magnetic field along the $\left[111\right]$ direction generates three inequivalent sites denoted 3a, 3b, and 3c (the red, green and blue circles), while in the field along the $\left[001\right]$ direction, there are two inequivalent sites denoted 3$\alpha$ and 3$\beta$. (the open and close circles).
}
\end{minipage} 
\end{figure}

\section{Knight shift and hyperfine coupling tensor}
The hyperfine field $\mathbf{H}_{\rm{hf}}$ acting on  $^{27}$Al nuclei produced by the Pr moment $\mathbf{m}$ can be written as, 
\begin{equation}
\mathbf{H}_{\rm{hf}}=\hat{A}_{hf} \mathbf{m},
\end{equation}
where $\hat{A}_{hf}$ is the hyperfine coupling tensor between  $\mathbf{m}$ and the nuclei. Because the crystal structure has the cubic symmetry, the magnetic susceptibility $\mathit{\chi}$ should be isotropic in the paramagnetic phase. Thus $\mathbf{m}$ is written by 
\begin{equation}
\mathbf{m}=\chi\mathbf{H}_{\rm{ext}}.
\end{equation}
Since $\left|\mathbf{H}_{\rm{ext}}\right|\gg\left|\mathbf{H}_{\rm{hf}}\right|$ in our case, only the component of $\mathbf{H}_{\rm{hf}}$ parallel to $\mathbf{H}_{\rm{ext}}$ contributes to the shift $K$. Therefore, $K$ is described as, 
\begin{equation}
K=\frac{\mathbf{H}_{\rm{ext}}\cdot\mathbf{H}_{\rm{hf}}}{\left|\mathbf{H}_{\rm{ext}}\right|^{2}}.
\end{equation}
From the equations (2), (3), and (4), we obtain
\begin{equation}
K=\frac{\mathbf{H}_{\rm{ext}}\cdot\left(\hat{A}_{hf} \mathbf{H}_{\rm{ext}}\right)}{\left|\mathbf{H}_{\rm{ext}}\right|^{2}} \chi \equiv a_{hf} \chi.
\end{equation}
The hyperfine coupling constant $a_{hf}$ is defined as the proportionality constant between $K$ and $\chi$. 

Figure 6 shows the $K-\chi$ plot above 6 K for the 1$\alpha$ site and all the five Al(3) sites. They all show good linear relation. From the slopes of these plots, the values of the hyperfine coupling constants $a_{hf}$ are determined as listed in table 2. We should remark here that the observed Knight shift includes contributions from the demagnetizing field and the Lorentz field, the latter being equal to $4 \pi M/3$ ($M$ is the magnetization per volume), in addition to the contribution from the hyperfine field. We have subtracted the contribution from demagnetizing field (estimated from the crystal shape and magnetization data) and the Lorentz field to obtain the values in table 2. The previous study [13] reported that $a_{hf} = 0.170$ T$/\mu_{B}$ for 3a sites, which is consistent with our result.

\begin{figure}
\begin{center}
\includegraphics[width=30pc]{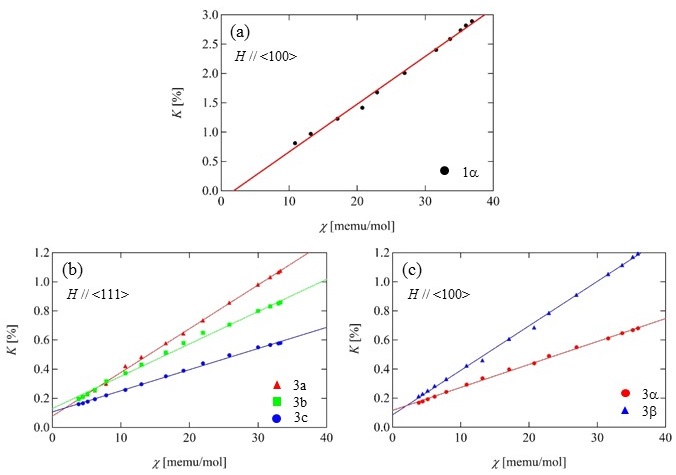}
\end{center}
\caption{\label{label}
$K$ vs $\chi$ plots of (a) the 1$\alpha$ site for $\mathbf{H}_{\rm{ext}} \parallel \left\langle 100\right\rangle $, (b) the 3a, 3b, and 3c sites for $\mathbf{H}_{\rm{ext}} \parallel \left\langle 111\right\rangle $, and (c) the 3a and 3b sites for $\mathbf{H}_{\rm{ext}} \parallel \left\langle 100\right\rangle $. 
}
\end{figure}

\begin{center}
\begin{table}[h]
\caption{hyperfine coupling constants for the 1$\alpha$ and all Al(3) sites in PrTi$_2$Al$_{20}$}
\footnotesize\rm
\centering
\begin{tabular}{@{}*{7}{c}}
\br
site&1$\alpha$&3a&3b\\
\mr
$a_{hf}$ [T/$\mu_{B}$]&$0.458\pm0.002$&$0.170\pm0.001$&$0.130\pm0.002$\\
\br
site&3c&3$\alpha$&3$\beta$\\
\mr
$a_{hf}$ [T/$\mu_{B}$]&$0.084\pm0.001$&$0.088\pm0.001$&$0.170\pm0.001$\\
\br
\end{tabular}
\end{table}
\end{center}

Based on these results and the crystal symmetry, we have determined all the components of the hyperfine coupling tensor $\hat{A}_{hf}$. Figure 7 shows a Pr atom at $\left[\frac{1}{8}\frac{1}{8}\frac{1}{8}\right]$ and Al(3) sites surrounding around the Pr atom. Here we explicitly consider the Al(3) site (the open circle) in figure 7, which is on the $\left(\bar{1}10\right)$ mirror plane. Because the $a$ and $b$ axes are equivalent for this site, the components of the hyperfine coupling tensor in the crystalline coordinate system can be written as

\begin{equation}
\hat{A}_{hf}=\left(\begin{array}{ccc}
A_{aa} & A_{ba} & A_{ca}\\
A_{ba} & A_{aa} & A_{ca}\\
A_{ca} & A_{ca} & A_{cc}
\end{array}\right) , 
\end{equation}
\begin{figure}
\begin{center}
\includegraphics[width=20pc]{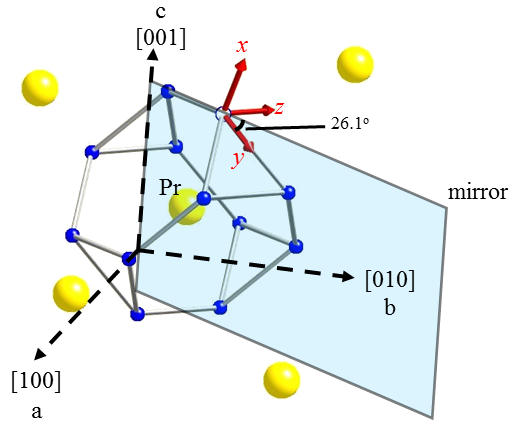}
\end{center}
\caption{
A part of the crystal structure showing the point symmetry of the Al(3) site identical to that shown in figure 6. The $x$, $y$, and $z$ axes represent the principal axes of the hyperfine coupling tensor between the Al(3) site (the open circle) and the Pr atom at the center of the cage, chosen to satisfying the condition $\left|A_{zz}\right|\geq\left|A_{yy}\right|\geq\left|A_{xx}\right|$. 
}
\end{figure}\parindent = 0pt
which includes four independent parameters, $A_{aa}$, $A_{cc}$, $A_{ba}$, and $A_{ca}$. The hyperfine coupling tensors at other Al(3) sites can be obtained by appropriate symmetry operations and their components are expressed as linear combinations of these parameters. Therefore, by using the equation (5), the hyperfine coupling constants $a_{\rm hf}$ at any site can be expressed as a linear combination of these four parameters. Since there are five experimental values of $a_{hf}$ listed in table 2, we have five linear equations with four parameters to be determined. We solved four out of the five equations to determine the values of $A_{aa}$, $A_{cc}$, $A_{ba}$, and $A_{ca}$ and used the remaining one equation for the consistency check. The results are satisfactory within the experimental error. We then diagonalized the $\hat{A}_{hf}$ tensor to obtain the principal values and the direction of principal axes. The results are summarized in table 3. Here $x$, $y$, and $z$ denote the principle axes chosen to satisfy the condition $\left|A_{zz}\right|\geq\left|A_{yy}\right|\geq\left|A_{xx}\right|$. As shown in figure 7, the $z$ axis is along the $\left[1\bar{1}0\right]$ direction and the $y$ axis is tilted from $\left[110\right]$ direction by $\beta=26.1\pm3.3$ degree. We also show in table 3 the components of the classical dipole coupling tensor $\hat{A}_{d}$. 
\begin{center}
\begin{table}[h]
\caption{The principle values of $\hat{A}_{hf}$ and $\hat{A}_{d}$}
\centering
\begin{tabular}{@{}*{7}{c}}
\br
&$A_{xx}$ [T/$\mu_{B}$]&$A_{yy}$ [T/$\mu_{B}$]&$A_{zz}$ [T/$\mu_{B}$]&$\beta$ [deg.]\\
\mr
$\hat{A}_{hf}$&$0.073\pm0.009$&$0.135\pm0.001$&$0.215\pm0.005$&$26.1\pm3.3$\\
$\hat{A}_{d}$&-0.035&0.045&-0.011&24.5\\
\br
\end{tabular}
\end{table}
\end{center}
The values of $\hat{A}_{hf}$ are several times larger than the values expected from the classical dipolar coupling mechanism, which confirms the dominance of the transferred hyperfine coupling mechanism. The transferred hyperfine coupling arises from the spin polarization of the conduction electrons at the Al site due to the Pr 4$f$ electrons. The values of $\hat{A}_{hf}$ thus can be a measure of the strength of the $c$-$f$ hybridization effects and the anisotropy of $\hat{A}_{hf}$ seem to be associated with the anisotropy of $c$-$f$ hybridization.

\section{Conclusion}
We have measured the $^{27}$Al-NMR spectra of a single crystal of PrTi$_2$Al$_{20}$. We have assigned NMR spectra to all Al sites forming the cage lattice and obtained the temperature dependences of the Knight shift at 1$\alpha$ and all Al(3) sites. The hyperfine coupling constants $a_{hf}$ were determined from the slopes of the $K$-$\chi$ plots and we obtained the hyperfine coupling tensor at the Al(3) site. This hyperfine tensor indicates that this system has a strong $c$-$f$ hybridization effect and the transferred hyperfine coupling is anisotropic. The results provide the quantitative basis to discuss the quadrupole order at low temperatures and effects of multipole fluctuations, which will be the subjects of future publications.  

\section{Acknowledgement}
The work was supported by Grant-in Aids for the Japan Society for the Promotion Science (JSPS) KAKENHI (B) (No. 25287083). T T was supported by the JSPS through the Program for Leading Graduate Schools (MERIT).

\smallskip
\end{document}